\documentstyle[preprint,aps]{revtex}
\tighten
\begin{document}
\draft
\title{Holographic Bound in Brans-Dicke Cosmology}
\preprint{UTEXAS-HEP-99-14}
\author{Yungui Gong} 
\address{Physics Department, University of Texas at Austin, Austin, 
Texas 78712}
\maketitle
\begin{abstract}
We apply the holographic principle to the Brans-Dicke cosmology.
We analyze the holographic bound in both the Jordan and Einstein
frames. The holographic bound is satisfied for both the $k=0$
and $k=-1$ universe, but it is violated for the $k=1$ matter
dominated universe.
\end{abstract}
\section{Introduction}

In black hole theory, we know that the total entropy of matter inside
a black hole cannot be greater than the Bekenstein-Hawking entropy,
which is $1/4$ of the area of the event horizon of the black hole
measured in Planck units \cite{benken}. The extension of this statement to more
general situations leads to the holographic principle \cite{holo}.
The most radical version of the holographic principle motivated by the Ads/CFT 
conjecture is that all the information about a physical system in a spatial
region is encoded in the boundary. The application of this idea to cosmology
was first considered by Fischler and Susskind \cite{fs}. For the universe,
it does not have a boundary, how can we apply the holographic principle
to cosmology? Fischler and Susskind answered this question by considering
a space inside the particle horizon. They proposed that the matter
entropy inside a spatial volume of particle horizon would not exceed
$1/4$ of the area of the particle horizon measured in Planck units.
They found that the flat universe and open universe obeyed this version
of holographic principle. However, closed universe violates this principle.
This may imply our universe is flat or open. On the other hand, this may
imply we need to revise the holographic principle somehow.
Easther and Lowe use the generalized second law of thermodynamics
to replace the holographic principle \cite{redl}. Bak and Rey \cite{dbsr}
considered apparent horizon instead of particle horizon to solve
the problem. In cosmology, there is a nature choice of length scale,
the Hubble distance, $H^{-1}$. $H^{-1}$ coincides with the particle
horizon and apparent horizon apart from an order 1 numerical factor
for the flat universe, but it becomes much larger than the apparent
horizon for the closed universe. So we know that the choice of
$H^{-1}$ as the horizon cannot solve the problem of violation of
the holographic principle in the closed universe.
The holographic principle in cosmology is also discussed in \cite{cosmo}
and \cite{bousso}.
Einstein's theory may not describe gravity at very high energy.
The simplest generalization of Einstein's theory is Brans-Dicke theory.
The recent interest in scalar-tensor theories of gravity arises from
the inflationary cosmology, supergravity and string theory. There
exists at least one scalar field, the dilaton field, in the low
energy effective bosonic string theory. Scalar degrees of freedoms
arise also upon compactification of higher dimensions.
In this paper, we apply the Fischler and Susskind proposal to the Brans-Dicke
cosmology in both the Jordan and Einstein frames.

\section{Brans-Dicke Cosmology in Jordan Frame}

The Brans-Dicke Lagrangian in Jordan frame is given by
\begin{equation}
\label{bdlagr}
{\cal L}_{BD}=-\sqrt{-\gamma}\left[\phi{\tilde  R}+
\omega\,\gamma^{\mu\nu}
{\partial_\mu\phi\partial_\nu\phi\over \phi}\right]-{\cal L}_m(\psi,\,
\gamma_{\mu\nu}).
\end{equation}
The above Lagrangian (\ref{bdlagr}) is conformal invariant under
the conformal transformations,
$$g_{\mu\nu}=\Omega^2\gamma_{\mu\nu},\quad \Omega=\phi^\lambda,~~
(\lambda\neq {1\over 2}),\quad \sigma=\phi^{1-2\lambda},
\quad {\bar \omega}={\omega-6\lambda(\lambda-1)\over (2\lambda-1)^2}.$$
For the case $\lambda=1/2$, we make the following transformations
\begin{equation}
\label{conformala}
g_{\mu\nu}=e^{\alpha\sigma}\gamma_{\mu\nu},
\end{equation}
\begin{equation}
\label{conformalb}
\phi={1\over 2\kappa^2}e^{\alpha\sigma},
\end{equation}
where $\kappa^2=8\pi G$, $\alpha=\beta\kappa$, and $\beta^2=2/(2\omega+3)$. 
Remember that the Jordan-Brans-Dicke Lagrangian is not
invariant under the above transformations (\ref{conformala})
and (\ref{conformalb}).
The homogeneous and isotropic
Friedman-Robertson-Walker (FRW) space-time metric is
\begin{equation}
\label{rwcosm}
ds^2=-dt^2+a^2(t)\left[{dr^2\over 1-k\,r^2}+r^2\,d\Omega\right],
\end{equation}
the above metric can be written as
\begin{equation}
ds^2=-dt^2 + a^2(t)\left[d\chi^2 + \Sigma^2 (d \theta^2 + \sin^2
\theta d\phi^2) \right],
\label{1.2b}
\end{equation}
where 
\begin{equation}
\Sigma=\left\{\begin{array}{lc}
\chi&k=0,\\
\sinh \chi&k=-1,\\
\sin \chi&k=1.
\end{array}\right.
\end{equation}

Based on the FRW metric and the perfect fluid
$T_m^{\mu\nu}=(\rho+ p)\,U^\mu\,U^\nu + p\,g^{\mu\nu}$
as the matter source,
we can get the evolution equations
of the universe from the action (\ref{bdlagr})
\begin{equation}
\label{jbd1}
H^2+{k\over a^2}+H{\dot{\phi}\over \phi}-{\omega\over 6}\left({\dot{\phi}
\over \phi}\right)^2={8\pi\over 3\phi}\rho,
\end{equation}
\begin{equation}
\label{jbd2}
\ddot{\phi}+3H\dot{\phi}=4\pi\beta^2(\rho-3p),
\end{equation}
\begin{equation}
\label{jbd3}
\dot{\rho}+3H(\rho+p)=0.
\end{equation}
If we are given a state equation for the matter $p=\gamma\rho$, then the
solution to Eq. (\ref{jbd3}) is
\begin{equation}
\label{jbd4}
\rho\,a^{3(\gamma+1)}=C_1.
\end{equation}

Most of the cosmological solutions in this paper were given in
\cite{dlp}.
For the case $k=0$, we can get the power-law
solutions to the Eqs. (\ref{jbd1}) and (\ref{jbd2})
with the help of Eq. (\ref{jbd4}),
\begin{equation}
\label{jbd5}
a(t)=a_0\,t^p,\quad \phi(t)=\phi_0\,t^q,
\end{equation}
where 
\begin{equation}
\label{jbd6}
p={2+2\omega(1-\gamma)\over 4+3\omega(1-\gamma^2)}, \quad
q={2(1-3\gamma)\over 4+3\omega(1-\gamma^2)},\quad -1\le \gamma <
1-{2\over 3+\sqrt{6}/\beta},
\end{equation}
$a_0$ and $\phi_0$ are integration constants,
and $[q(q-1)+3pq]\phi_0=4\pi\beta^2(1-3\gamma)C_1 a_0^{-3(\gamma+1)}$.
The solution for $\gamma=1/3$ is very special because the scalar field does
not evolve. We will take care of this case later.
The particle horizon is 
\begin{equation}
\label{jbd7}
r_H=\int_0^t {d{\tilde t}\over a({\tilde t})}={4+3\omega(1-\gamma^2)\over a_0[2+\omega
(1-\gamma)(1+3\gamma)]}t^{1-p}.
\end{equation}
Therefore, the ratio between the entropy inside the particle horizon
and the area of the horizon is
\begin{equation}
\label{jbd8}
{S\over GA/4}={4 \over 3G}\epsilon {r_H\over a^2}
={4\epsilon \over 3G}\,
{4+3\omega(1-\gamma^2)\over a_0^3[2+\omega
(1-\gamma)(1+3\gamma)]}t^{1-3p},
\end{equation}
where $\epsilon$ is the constant comoving entropy density,
and $1-3p=-[2+3\omega(1-\gamma)^2]/[4+3\omega(1-\gamma^2)]$.
The holographic bound is satisfied for $\gamma$ in the range
given by Eq. (\ref{jbd6}) if the above 
ratio is not greater than 1 initially.

For the case $k=\pm 1$, we do not have a general solution for all
values of $\gamma$,
so we consider two special cases:
the matter dominated universe with $\gamma=0$ and the
radiation dominated universe with $\gamma=1/3$. It is
convenient to use the cosmic time $d\eta=dt/a(t)$.

For $\gamma=1/3$, we can solve Eq. (\ref{jbd2}) to get
\begin{equation}
\label{jbd21}
a^3\dot{\phi}=C_2,
\end{equation}
where $C_2\neq 0$ is an integration constant. 

\begin{itemize}
\item[(i)] $k=1$, the solutions are
\begin{equation}
\label{jbd22}
\phi(\eta)=\phi_0\left[{4\pi C_1 \tan(\eta+\eta_0)/3+\sqrt{16\pi^2 C_1^2/9+C_2^2/6\beta^2}
-\sqrt{C_2^2/6\beta^2}\over
4\pi C_1 \tan(\eta+\eta_0)/3+\sqrt{16\pi^2 C_1^2/9+C_2^2/6\beta^2}
+\sqrt{C_2^2/6\beta^2}}\right]^{\sqrt{3\beta^2/2}},
\end{equation}
\begin{equation}
\label{jbd23}
a^2(\eta)\phi(\eta)={4\pi C_1\over 3}+\sqrt{{16\pi^2 C_1^2\over 9}+
{C_2^2\over 6\beta^2}}\,\sin[2(\eta+\eta_0)],
\end{equation}
where $\eta_0$ is an integration constant. The entropy
to area ratio is
\begin{equation}
\label{jbd24}
{S\over GA/4}={\epsilon(2\eta-\sin 2\eta)\phi(\eta)\over 
G\sin^2\eta \{4\pi C_1/3+\sqrt{64\pi^2 C_1^2/9+
2C_2^2/3\beta^2}\,\sin[2(\eta+\eta_0)]\}}.
\end{equation}
Note that $0\le 2(\eta+\eta_0)\le \pi$, so we
see that the holographic bound can be satisfied if
it is satisfied initially.

\item[(ii)] $k=-1$ and $C_2^2<32\pi^2\beta^2 C_1^2/3$, we have solutions
\begin{equation}
\label{jbd28}
a^2(\eta)\phi(\eta)=-{4\pi C_1\over 3}+{1\over 16}e^{2(\eta+\eta_0)}+
\left({64\pi^2 C_1^2\over 9}-{2 C_2^2\over 3\beta^2}\right)e^{-2(\eta+\eta_0)},
\end{equation}
\begin{equation}
\label{jbd29}
\phi=\phi_0\left[
{-(4\pi C_1/3+b)\tanh(\eta+\eta_0)
-c+C_2/\beta\sqrt{6}\over
-(4\pi C_1/3+b)\tanh(\eta+\eta_0)
-c-C_2/\beta\sqrt{6}}\right]^{\sqrt{3\beta^2/2}},
\end{equation}
where $b=1/16+64\pi^2 C_1^2/9-2 C_2^2/3\beta^2$ and
$c=1/16-64\pi^2 C_1^2/9+2 C_2^2/3\beta^2$.
The Brans-Dicke scalar field changes very slowly compared to
the scale factor. Therefore the holographic bound
\begin{equation}
\label{jbd30}
{S\over GA/4}={\epsilon(\sinh 2\eta-2\eta)\phi\over
G\sinh^2 \eta [-4\pi C_1/3+e^{2(\eta+\eta_0)}/16+
(64\pi^2 C_1^2/9-2 C_2^2/3\beta^2)e^{-2(\eta+\eta_0)}]}\le 1,
\end{equation}
will be satisfied if it is satisfied initially.

\item[(iii)] $k=0$, the solutions are
\begin{equation}
\label{jbd25}
a^2(\eta)\phi(\eta)={8\pi C_1\over 3} (\eta+\eta_0)^2-{C_2^2\over
16\pi\beta^2 C_1},
\end{equation}
\begin{equation}
\label{jbd26}
\phi(\eta)=\phi_0\left[{\eta+\eta_0-\sqrt{6}\, C_2/16\pi\beta C_1\over
\eta+\eta_0+\sqrt{6}\,C_2/16\pi\beta C_1}\right]^{\sqrt{3\beta^2/2}}.
\end{equation}
The Brans-Dicke
scalar field $\phi$ slowly increases up to $\phi_0$ as the universe expands.
The holographic bound
\begin{equation}
\label{jbd27}
{S\over GA/4}={4\epsilon\over 3G}\,{\eta\phi(\eta)\over
8\pi C_1(\eta+\eta_0)^2/3-C_2^2/16\pi\beta^2 C_1}\le 1
\end{equation}
can be satisfied if it is satisfied initially.
\end{itemize}

For $\gamma=0$, the solutions are:
\begin{equation}
\label{jbd9}
a(\eta)=a_0 e^{b\eta}, \qquad \phi=\phi_0 e^{-b\eta},
\end{equation}
where $b^2=-2k/(2+\omega)$ and $4\pi\beta^2 C_1=
-a_0\phi_0 b^2$.
\begin{itemize}
\item[a.] $k=-1$ and $-2<\omega<-3/2$,
the above solutions (\ref{jbd9}) are exponential expansion in the cosmic
time $\eta$ or linear expansion in the coordinate time $t$. The entropy
to area ration is
\begin{equation}
\label{jbd10}
{S\over GA/4}={\epsilon(\sinh 2\eta -2\eta)\over G a_0^2 e^{2b\eta}\sinh^2\eta}.
\end{equation}
So the holographic bound can be satisfied for $-2<\omega<-3/2$
if it is satisfied initially.

\item[b.] $k=1$ and $\omega<-2$, the solutions (\ref{jbd9}) are linear
in the coordinate time $t$. The entropy to area ratio is
\begin{equation}
\label{jbd11}
{S\over GA/4}={\epsilon (2\eta-\sin 2\eta)\over G a_0^2 e^{2b\eta} \sin^2\eta}.
\end{equation}
It is obvious that the holographic bound can be violated when $\eta=n\pi$
for any integer $n$.
\end{itemize}
In fact, the current experimental constraint on $\omega$ is $\omega>500$
or $\beta^2<0.002$.
The solutions (\ref{jbd9}) may not be physical. However, the low energy
effective theory of the string theory can lead to $\omega=-1$, we
may need to explore the possibility of negative $\omega$. 
For positive $\omega$, we need to solve the equations numerically.
When $\omega \rightarrow \infty$
and at late times,
the Brans-Dicke cosmological solutions become general relativistic solutions. 

\section{Brans-Dicke Cosmology in Einstein Frame}

The Brans-Dicke Lagrangian in Einstein frame is obtained by the conformal transformations
(\ref{conformala}) and (\ref{conformalb}),
\begin{equation}
\label{1}
{\cal L}= \sqrt{-g} \left[-\frac{1}{2\kappa^2} R
-\frac{1}{2}g^{\mu\nu}\partial_\mu \sigma \partial_\nu \sigma\right]
-{\cal L}_{m}(\psi, e^{-\alpha\sigma}g_{\mu\nu}).
\end{equation}

The perfect fluid becomes
$T_m^{\mu\nu}=e^{-2\,a\sigma}[(\rho+ p)\,U^\mu\,U^\nu + p\,g^{\mu\nu}]$.
From the FRW metric in the Einstein frame,
we can get the evolution equations
of the universe from the action (\ref{1})
\begin{equation}
\label{3.1a}
H^2+{k \over a^2}={\kappa^2 \over 3}\left({\frac 1 2}\dot{\sigma}^2
+e^{-2\alpha\sigma}\rho\right),
\end{equation}
\begin{equation}
\label{3.1b}
\ddot{\sigma}+3H\dot{\sigma}={\frac 1  2}\alpha e^{-2\alpha\sigma}(\rho-3p),
\end{equation}
\begin{equation}
\label{3.1c}
\dot{\rho} +3H(\rho + p)={\frac 3 2}\alpha\dot{\sigma}(\rho +p).
\end{equation}
With $p=\gamma\rho$, the solution to Eq. (\ref{3.1c}) is
\begin{equation}
\label{3.1ci}
\rho\,a^{3(\gamma+1)}\,e^{-3\alpha(\gamma+1)\sigma/2}=C_3,
\end{equation}
where $C_3$ is a constant of integration.
For the flat universe $k=0$,
combining Eqs. (\ref{3.1a}), (\ref{3.1b}) and (\ref{3.1c}), we have
\begin{equation}
\label{3.1i}
a\,e^{-\alpha(1-\gamma)\sigma/\beta^2(1-3\gamma)}=C_4,
\end{equation}
where $C_4$ is an integration constant and
the above equation is valid for $-1\le \gamma <1-2/(3+\sqrt{6}/
\beta)$ and $\gamma\neq 1/3$.
To obtain the above solution, we assume that $\dot{\sigma}a^3\rightarrow 0$
and $\dot{a}a^2\rightarrow 0$ when $a\rightarrow 0$. From Eqs. (\ref{3.1a}),
(\ref{3.1ci}) and (\ref{3.1i}), we get
\begin{equation}
\label{ebd1}
H^2={2\kappa^2(1-\gamma)^2 C_3 C_4^{\beta^2(1-3\gamma)^2/2(1-\gamma)}\over 
6(1-\gamma)^2-\beta^2(1-3\gamma)^2}a^{-[6(1-\gamma^2)+\beta^2(1-3\gamma)^2]
/2(1-\gamma)}.
\end{equation}
The particle horizon is
\begin{equation}
\label{ebd2}
r_H=\int_0^a {d{\tilde a}\over {\tilde a}^2 H}=B
a^{[2(1-\gamma)(1+3\gamma)+\beta^2(1-3\gamma)^2]/4(1-\gamma)},
\end{equation}
where 
$$B={4C_4^{-\beta^2(3\gamma-1)^2/4(1-\gamma)}\sqrt{6(1-\gamma)^2
-\beta^2(1-3\gamma)^2}\over [2(1-\gamma)(1+3\gamma)+\beta^2(3\gamma-1)^2]
\sqrt{2\kappa^2 C_3}}$$ 
is a constant coefficient.
The entropy to area ratio is
\begin{equation}
\label{ebd3}
{S\over GA/4}={4\epsilon B\over 3G} a^{-[6(1-\gamma)^2-\beta^2(1-3\gamma)^2]/4(1-\gamma)}.
\end{equation}

For $\gamma=1$, we find that
$$H a^3=C_6,$$
where $C_6$ is an integration constant. The entropy to area ratio
$${S\over GA/4}={2\epsilon\over 3 G C_6},$$
is also a constant.

For $\gamma=1/3$, we have
\begin{equation}
\label{ebd4}
a^3\dot{\sigma}=C_5,
\end{equation}
where $C_5\neq 0$ is an integration constant.
\begin{enumerate}
\item $k=0$, the entropy to area ratio is
\begin{equation}
\label{ebd5}
{S\over GA/4}={4\epsilon \over G \kappa C_3}
{\sqrt{C_3a^2+C_5^2/2}-\sqrt{C_5^2/2}\over \sqrt{3} a^2}.
\end{equation}
Therefore, from Eqs. (\ref{ebd3}) and (\ref{ebd5}), we see
that the holographic principle is satisfied for 
$-1\le \gamma <1-2/(3+\sqrt{6}/\beta)$ 
provided that it is satisfied initially.

\item $k=-1$ and $\kappa^2 C_3^2\ge 6 C_5^2$,
we have
\begin{equation}
\label{ebd6}
e^{2\chi_H}={2\sqrt{a^4+\kappa^2 C_3^2 a^2/3+\kappa^2 C_5^2/6}+
2a^2+\kappa^2 C_3/3\over 2\sqrt{\kappa^2 C_5^2/6}+\kappa^2 C_3/3}.
\end{equation}
The entropy to area ratio is
\begin{equation}
\label{ebd7}
{S\over GA/4}={\epsilon(\sinh 2\chi_H-2\chi_H)\over
G a^2\sinh^2\chi_H}.
\end{equation}
As $a$ increases, $4S/GA$ decreases. The holographic bound
is satisfied if it is satisfied initially.

\item $k=1$, we have
\begin{equation}
\label{ebd8}
2\chi_H=\arcsin {\kappa C_3\over \sqrt{\kappa^2 C_3^2+6 C_5^2}}
+\arcsin {6a^2-\kappa^2 C_3\over \sqrt{\kappa^4 C_3^2+6 \kappa^2 C_5^2}}.
\end{equation}
The holographic bound
\begin{equation}
\label{ebd9}
{S\over GA/4}={\epsilon(2\chi_H-\sin 2\chi_H)\over
G a^2\sin^2\chi_H}\le 1
\end{equation}
is satisfied if it is satisfied initially.
\end{enumerate}

For $\gamma=0$ and $k^2=1$, we do not have any analytical solution. 
We need to solve the problem numerically.

\section{Conclusions}

We analyze the holographic principle in Brans-Dicke theory. For the
flat universe, we find that the holographic bound can be satisfied
for any matter with $-1\le \gamma <1-2/(3+\sqrt{6}/\beta)$. For the
universe with $k^2=1$, we do not have general analytical solutions
for all values of $\gamma$. In particular, we do not have analytical
solution for the matter dominated $k^2=1$ universe.  We know
that in standard Friedman cosmology, the holographic principle is violated
for the closed matter dominated universe near the maximal expansion. 
To check the holographic bound
for the $k=1$ matter dominated Brans-Dicke cosmological model, we need
to do numerical calculation. However, the numerical results in \cite{bdnum}
tell us that the expansion rate in Brans-Dicke models are slower
than those in Friedman models. At large times, the difference
becomes negligible. Therefore we expect that the holographic
bound is also violated for the $k=1$ matter dominated universe
in Brans-Dicke cosmology. 

More recently, Bousso \cite{bousso} consider the holographic bound
in regions
generated by null geodesics. He gave a prescription to select
light sheets which are hypersurfaces generated by surfaces orthogonal
to null geodesics with non-positive expansion. This covariant
entropy bound can be hold in general space times\footnote{The author
thanks Raphael Bousso for the references.}. I believe
that the covariant entropy bound is also satisfied for
Brans-Dicke cosmological models. To defend this belief, we
need to do numerical calculation.

\end{document}